\documentclass[preprints,review,accept,moreauthors,pdftex,10pt,a4paper,amsmath, amssymb]{mdpi} 
\usepackage {bm}
\usepackage{color}
\firstpage{1} 
\pubvolume{xx}
\issuenum{1}
\articlenumber{5}
\pubyear{2018}
\copyrightyear{2018}
\history{}
\def\kF{k_{\text{F}}}
\def\vF{v_{\text{F}}}
\def\NF{N_{\text{F}}}

\def\epsilonF{\epsilon_{\text F}}

\def\sgn{{\text{sgn\,}}}
\def\be{\begin{equation}}
\def\ee{\end{equation}}
\def\bea{\begin{eqnarray}}
\def\eea{\end{eqnarray}}
\def\bse{\begin{subequations}}
\def\ese{\end{subequations}}

\def\agt{\substack{ >\\ \sim}}
\def\alt{\substack{<\\ \sim}}
\Title{Anomalous Transport Behavior in Quantum Magnets}

\Author{Dietrich Belitz $^{1}$*\ %\orcidA{} 
                and Theodore R. Kirkpatrick $^{2}$}

\AuthorNames{Dietrich Belitz and Theodore R. Kirkpatrick}

\address{%
$^{1}$ \quad Dept. of Physics, Institute of Theoretical Science, and Materials Science Institute, Univ. of Oregon, Eugene, OR 97403, USA; dbelitz@uoregon.edu\\
$^{2}$ \quad Institute for Physical Science and Technology, Univ. of Maryland, College Park, MD 20742, USA; tedkirkp@umd.edu}

\corres{Correspondence: dbelitz@uoregon.edu; Tel.: +1-541-346-4738}

\abstract{Transport behavior characterized by a low-temperature electrical resistivity that displays a 
power-law behavior $\rho(T\to 0) \propto T^s$, with an exponent $s<2$, is commonly observed in magnetic
materials in both the magnetic and nonmagnetic phases. We give a pedagogical overview of this phenomenon 
that summarizes both the experimental situation and the state of its theoretical understanding. We also put it in 
context by drawing parallels with unusual power-law transport behavior in other systems.}


\begin{document}
%%%%%%%%%%%%%%%%%%%%%%%%%%%%%%%%%%%%%%%%%%
%% Only for the journal Gels: Please place the Experimental Section after the Conclusions

%%%%%%%%%%%%%%%%%%%%%%%%%%%%%%%%%%%%%%%%%%
%\setcounter{section}{-1} %% Remove this when starting to work on the template.
%\section{How to Use this Template}
%The template details the sections that can be used in a manuscript. Note that the order and names of article sections may differ from the requirements of the journal (e.g. the positioning of the Materials and Methods section). Please check the instructions for authors page of the journal to verify the correct order and names. For any questions, please contact the editorial office of the journal or support@mdpi.com. For LaTeX related questions please contact Janine Daum at latex-support@mdpi.com.

%The order of the section titles is: Introduction, Materials and Methods, Results, Discussion, Conclusions for these journals: aerospace,algorithms,antibodies,antioxidants,atmosphere,axioms,biomedicines,carbon,crystals,designs,diagnostics,environments,fermentation,fluids,forests,fractalfract,informatics,information,inventions,jfmk,jrfm,lubricants,neonatalscreening,neuroglia,particles,pharmaceutics,polymers,processes,technologies,viruses,vision

\section{Introduction}
\label{sec:1}

\begin{figure}[t]
\centering
\includegraphics[width=12 cm]{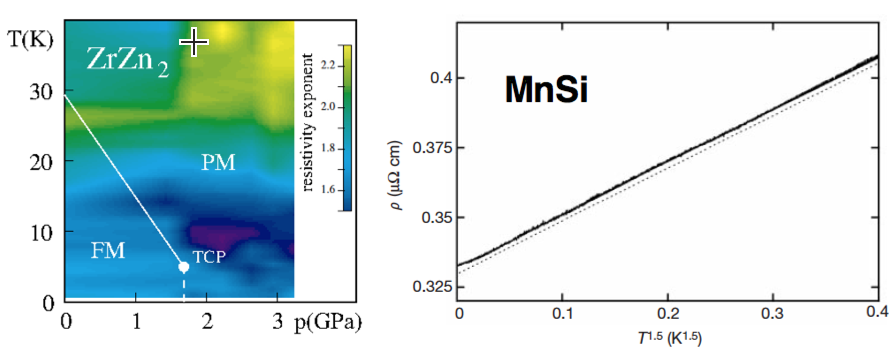}
\caption{{\it Left panel:} Observed temperature-pressure phase diagram of ZrZn$_2$, with the false colors indicating the value of the resistivity exponent $s$. 
               The white lines represent lines of second order (solid) and first order (dashed) transitions between the ferromagnetic (FM) and
               paramagnetic (PM) phases, and TCP denotes the tricritical point where the order of the transition changes. After Fig.~2 in
               Ref.~\cite{Takashima_et_al_2007}; this version taken from Ref.~\cite{Kirkpatrick_Belitz_2018}. {\it Right panel:} Measured resistivity of MnSi in the
               nonmagnetic phase. After Ref.~\cite{Pfleiderer_Julian_Lonzarich_2001}.}
\label{fig:1}
\end{figure}

Simple metals are characterized, {\it inter alia}, by a low-temperature ($T$) behavior of the electrical resistivity $\rho$ given by a power-law $\delta\rho(T\to 0) \propto T^2$ \cite{Wilson_1954,
Kittel_2005}, with $\delta\rho = \rho - \rho_0$ the temperature-dependent part of the resistivity and $\rho_0$ the residual resistivity. This is often considered one of the hallmarks of a Fermi 
liquid , and a stronger $T$-dependence of the form 
\begin{equation}
\delta\rho(T\to 0) =A_s T^s
\label{eq:1}
\end{equation}
with an exponent $s<2$ is often referred to as ``non-Fermi-liquid'' (NFL) (transport) behavior, although this designation requires some qualification, as we will discuss in Sec.~\ref{sec:5}. 
A prominent example is the linear $T$-dependence of the
resistivity in the normal phase of hole-doped high-T$_c$ superconductors near optimal doping \cite{Gurvitch_Fiory_1989, Takagi_et_al_1992}. 
Examples in other systems are provided by various ferromagnets with a low
Curie temperature. Sato observed a behavior given by Eq.~(\ref{eq:1}) with $s\approx 1.50 - 1.65$ in Pd-doped Ni$_3$Al \cite{Sato_1975}. Very similar behavior was found in
pressure-tuned Ni$_3$Al \cite{Niklowitz_et_al_2005}, as well as in pressure-tuned ZrZn$_2$ \cite{Takashima_et_al_2007}. Another example is provided by the
helical magnet MnSi, which shows a very clean $s=3/2$ behavior in a temperature range from a few mK to several K \cite{Pfleiderer_Julian_Lonzarich_2001}.
The measured resistivities of ZrZn$_2$ and MnSi are shown in Fig.~\ref{fig:1} as representative examples. We note that the anomalous transport behavior in ZrZn$_2$ is
observed in both the ordered and the disordered phases, whereas in MnSi it shows only in the disordered phase. In the helically ordered phase of MnSi one observes
$\delta\rho \propto T^2$, albeit with a large prefactor $A_2$, an observation we will come back to.
\vskip -0cm
\vskip 0cm

Surprisingly, this anomalous transport behavior is far from being completely understood, despite having been observed for many decades in many different
materials. In this paper we provide a pedagogical overview of this problem and the solutions that have been proposed.

%%%%%%%%%%%%%%%%%%%%%%%%%%%%%%%%%%%%%%%%%%%%%%%%%%%%%%%%%%%%%%%%%%%%% 
\section{Soft modes as the origin of power-law relaxation rates}
\label{sec:2}
 
It is intuitively plausible that any power-law behavior of relaxation rates, including those that determine transport coefficients, requires the scattering of conduction electrons by soft or
massless excitations, i.e., excitations whose characteristic frequency vanishes in the limit of vanishing wave number: A gapped excitation, whose frequency remains nonzero in this
limit, will get frozen out at temperatures small compared to the gap and produce an exponentially
small relaxation rate. This can be demonstrated explicitly by means of some very simple and general arguments.
 
As a very simple schematic example, consider noninteracting electrons described by an action $S_0[{\bar\psi},\psi]$. $\bar\psi(x)$ and $\psi(x)$ are
fermionic fields, $x\equiv({\vec x},\tau)$ comprises the real-space position $\vec x$ and the imaginary-time variable $\tau$, and we suppress discrete
degrees of freedom such as spin, band indices, etc., in our notation. Let the single-electron energy-momentum relation be $\epsilon_{\vec k}$, and 
denote the chemical potential by $\mu$. The Fermi surface is then characterized by $\xi_{\vec k} \equiv \epsilon_{\vec k} - \mu = 0$. Consider a generalized electron
density $n(x) = \bar\psi(x)\psi(x)$, its fluctuations $\delta n(x) = n(x) - \langle n(x)\rangle$, and denote its Fourier transform by $n(k)$, with $k\equiv(\vec k,\omega_n)$ 
a 4-vector that comprises a wave vector $\vec k$ and a fermionic Matsubara frequency $\omega_n$. Examples of $n(x)$ are the number density, the spin
density, or any other moment of a general phase-space density. In addition, let $\delta N(x)$ be a non-electronic density fluctuation that
is governed by a Gaussian action
\begin{equation}
S_{\text{fluct}}[\delta N] = \frac{-1}{2} \int dx\,dy\,\delta N(x)\,\chi^{-1}(x-y)\,\delta N(y)
\label{eq:2}
\end{equation}
with $\chi$ the physical susceptibility appropriate for the fluctuations $\delta N$, and couples to the electronic density via a short-ranged interaction potential $v$:
\begin{equation}
S_{\text{coup}} = \int dx\,dy\,\delta N(x)\,\delta(\tau_x - \tau_y)\,v({\vec x}-{\vec y})\,\delta n(y)\ .
\label{eq:3}
\end{equation}
An example of $\delta N$ are ionic density fluctuations, in which case $n$ is the electronic number density, $v$ is the screened Coulomb interaction, and
$S_{\text{coup}}$ describes the electron-phonon coupling. If we integrate out the fluctuations $\delta N$ we obtain an effective electronic action
\bse
\label{eqs:4}
\be
S_{\text{eff}}[\bar\psi,\psi] = S_0[\bar\psi,\psi] + \frac{1}{2}\int_k \delta n(k)\,V(k)\,\delta n(-k)
\label{eq:4a}
\end{equation}
with an effective potential $V(k) = \left(v({\vec k})\right)^2 \chi(k)$. Since the potential $v$ is short ranged we can, for the purpose of studying
long-wavelength effects, replace this expression by
\be
V(k) = C\, \chi(k)
\label{eq:4b}
\ee
\ese
with $C = v^2(\vec k=0)$ a coupling constant. 
The integration measures in Eqs.~(\ref{eq:2},\ref{eq:3}) and (\ref{eq:4a}), respectively, are $\int dx \equiv \int_V d{\vec x} \int_0^{1/T} d\tau$ and
$\int_k = (1/V)\sum_{\vec k} T\sum_{\omega_n}$, with $V$ the system volume. We use units such that the Boltzmann constant $k_{\text B}=1$.

The effective electron-electron interaction described by the potential $V$ can be interpreted as an exchange of $\delta N$ fluctuations by the electrons. 
It leads to an electron-self energy that is given, in Hartree-Fock approximation, by
\begin{equation}
\Sigma(p) = \int_k V(k)\,G(p-k)
\label{eq:5}
\end{equation}
The single-particle relaxation rate $1/\tau_{\text{sp}}$, i.e., the inverse quasiparticle lifetime due to the effective interaction, averaged over the Fermi surface, is given by
\bea
\frac{1}{2\tau_{\text{sp}}} &=& \frac{-1}{\NF V}\sum_{\vec p} \delta(\xi_{\vec p})\,\Sigma''({\vec p},0)
\nonumber\\
            &=& 2\NF \int_{-\infty}^{\infty} du\,{\bar V}''(u)\,\frac{1}{\sinh(u/T)}
\label{eq:6}
\eea
Here $\NF$ is the electronic density of states at the Fermi surface, $\Sigma''({\vec p},\omega) = \text{Im} \Sigma({\vec p},i\omega_n\to \omega + i0)$
is the spectrum of the self energy, and
\be
{\bar V}''(u) = \frac{1}{(\NF V)^2} \sum_{{\vec k},{\vec p}} \delta(\xi_{\vec k}) \delta(\xi_{\vec p})\,V''({\vec k}-{\vec p},u)
\label{eq:7}
\ee
is the spectrum of the effective potential averaged over the Fermi surface. For simplicity, we ignore the splitting of the Fermi surface in
magnets for the time being. We will add this feature, and several important others, in Sec.~\ref{sec:4}. 

\subsection{Power-law relaxation rates from exchange of particles}
\label{subsec:2.1}

To specify the effective potential $V$, consider a particle-like excitation with a wave-number dependent resonance frequency
$\omega_0({\vec k}\to 0) = c\,\vert{\vec k}\vert^n$, in which case the spectrum of the susceptibility $\chi$ has the form
\be
\chi''({\vec k},u) \propto \vert u\vert^m\,\sgn(u)\,\delta(u^2-\omega_0^2({\vec k}))
\label{eq:8}
\ee
Here $c$ is a stiffness coefficient, we ignore a prefactor that we absorb into the coupling constant $C$, and we neglect any damping of the
excitation. The values of the exponents $n$ and $m$ depend on the nature of the particles, we will see examples below.

Performing the wave-number integrals in Eq.~(\ref{eq:7}) we find
\be
{\bar V}''(u) \propto \vert u\vert^{m+(d-1-2n)/n} \sgn(u)
\label{eq:9}
\ee
Via Eq.~(\ref{eq:6}) this leads to 
\be
1/\tau_{\text{sp}} \propto T^{m+(d-1-n)/n}
\label{eq:10}
\ee
with $d$ the spatial dimensionality. 

These simple considerations illustrate a basic point: The power-law behavior of $1/\tau_{\text{sp}}$ hinges on the resonance frequency
$\omega_0$ scaling as a power of $\vert{\vec k}\vert$ for ${\vec k}\to 0$. This is the defining property of a mode that is soft,
or gapless, or massless. 

As a well-known example, consider longitudinal phonons. In this case, $\delta n$ and $\delta N$ are the electronic and ionic 
number density fluctuations, respectively, and $v$ is a screened Coulomb interaction. The susceptibility $\chi$ has the same form
as in a classical fluid and is characterized by $m=2$ and $n=1$ \cite{Forster_1975}. We thus have $1/\tau_{\text{sp}} \propto T^d$. In
$d=3$ this is the well-known $T^3$ law for the single-particle relaxation rate due to phonons \cite{Ziman_1960}.

We note that we have made several simplifying assumptions so far, in addition to the assumption of a single Fermi surface
mentioned above. First, we have considered only the single-particle relaxation rate, rather than the more complicated transport rate 
$1/\tau_{\text{tr}}$ which determines the electrical resistivity. (The single-particle rate does, however, have the same $T$-dependence
as the thermal resistivity, at least at the level of the Boltzmann equation \cite{Wilson_1954}.) Second, we have assumed an isotropic 
resonance frequency that depends only on the magnitude
of the wave vector. Third, we have ignored the effects of quenched disorder, which is always present in real materials, if
possibly only very weakly. Relaxing this constraints is important in order to understand the experimental results we are 
interested in; we will discuss this in Sec.~\ref{sec:4}.

\subsection{Power-law relaxation rates from exchange of unparticles}
\label{subsec:2.2}

Another possibility is the exchange of fluctuations that are described by a continuous spectrum that is scale invariant
but lacks the resonance peak characteristic of particles:
\be
\chi''({\vec k},u) \propto u^m/\vert{\vec k}\vert^n
\label{eq:11}
\ee
Spectra of this type are familiar from condensed-matter physics; the most common example is the Lindhard function \cite{Pines_Nozieres_1989}. In a 
particle-physics context such excitations have been dubbed `unparticles' \cite{Georgi_2007}. The wave-vector integrals in Eq.~(\ref{eq:7}) then simply 
lead to a prefactor, and the single-particle and transport rates are the same except for a prefactor of $O(1)$, $1/\tau_{\text{sp}} \approx 1/\tau_{\text{tr}} \equiv 1/\tau$. 
The temperature dependence of either relaxation rate is determined by the exponent $m$ only, and we have
\be
1/\tau \propto T^{m+1}
\label{eq:12}
\ee

An obvious example is the case of Coulomb scattering. In this case $\delta N$ and $\delta n$ both represent electronic
number-density fluctuations that interact via a screened Coulomb interaction. $\chi''$ then is the spectrum of the Lindhard function,
and hence $m=n=1$, which leads to $1/\tau\propto T^2$. We note that at the level of quantum electrodynamics, the objects exchanged by the
electrons in this example are of course particles, viz., virtual photons. However, at the level of an effective low-energy theory where the
microscopic details have been integrated out, the effects of this exchange manifest themselves in the form of a continuous spectrum, viz.,
the dynamically screened Coulomb interaction. 

For later reference we restore the prefactors, which leads to the familiar result for the Coulomb scattering rate
\be
1/\tau = \pi T^2/2\epsilonF
\label{eq:13}
\ee
with $\epsilonF$ the Fermi energy. The above derivation is similar in spirit to the one in Ref.~\cite{Anderson_1984}. It is remarkable that the 
argument of interacting density fluctuations still works if the fluctuation $\delta N$ that interacts with the electronic fluctuation $\delta n$ is itself
an electronic density fluctuation created by all the other electrons. This aspect was stressed in Ref.~\cite{Bharadwaj_Belitz_Kirkpatrick_2014}.

%%%%%%%%%%%%%%%%%%%%%%%%%%%%%%%%%%%%%%%%%%%%%%%%%%%%%%%%%%%%%%%%%%%%%%% 
\section{Experimental results}
\label{sec:3}
 
For a classification of experimental results that show anomalous transport behavior it is crucial to distinguish between two different cases.
In the first case, the anomalous behavior is observed only in a narrow region of the phase diagram, usually in the vicinity of a known or
suspected critical point. Its observation thus requires fine tuning. An example is the $T^{3/2}$ resistivity combined with a logarithmic
$T$-dependence of the specific-heat coefficient observed near a probable quantum critical point in NbFe$_2$ \cite{Brando_et_al_2008}.
In the second case, the anomalous behavior is generic in the sense that
it is observed in large regions of the phase diagram. This distinction is crucial, since critical points necessarily lead to critical fluctuations
that can serve as the scale invariant excitations underlying the mechanism discussed in Sec.~\ref{sec:2}. Another important distinction is
between clean systems that contain no or very little quenched disorder, and disordered ones. This is because quenched disorder leads
to diffusive electron dynamics that can lead to anomalous transport behavior via well-known mechanisms \cite{Lee_Ramakrishnan_1985,
Belitz_Kirkpatrick_1994}. The anomalous transport behavior that is hardest to understand thus occurs in systems that are clean, as evidenced by
a small residual electric resistivity $\rho_0$ or a large mean-free path, and show generic anomalous behavior that does not require fine
tuning.

Two materials that fall into the latter category are the ferromagnet ZrZn$_2$, and  the helimagnet MnSi. The cleanest samples for either
systems have a $\rho_0 \approx 0.3\mu\Omega$cm, and the transition from the magnetic to the nonmagnetic phase can be triggered
by applying hydrostatic pressure. The magnetic quantum phase transitions are well established to be first order 
\cite{Uhlarz_Pfleiderer_Hayden_2004, Pfleiderer_et_al_1997}, so critical fluctuations are not a viable candidate for explaining the observed 
transport anomalies. See Ref.~\cite{Brando_et_al_2016} and references therein for a review of the magnetic properties of these materials. 

The phase diagram of ZrZn$_2$ is shown in Fig.~\ref{fig:1}. The resistivity exponent $s$, determined by the slope of a log-log plot of the electrical 
resistivity, is less than $2$ in a large part of the phase diagram, ranging from ambient pressure to twice the critical pressure, and from
the lowest temperatures achievable to about 20K. The smallest exponent $s\approx 1.5$ was found in a temperature region around
10K in the paramagnetic phase \cite{Takashima_et_al_2007}. Data at ambient pressure have been fitted to Eq.~(\ref{eq:1}) with $s=5/3$ for samples with
residual resistivities  between $0.3\mu\Omega$cm and $6.4\mu\Omega$cm, while a magnetic field of 9T restores a $T^2$ behavior
\cite{Sutherland_et_al_2012}. The respective prefactors are $A_{5/3} \approx 0.021\mu\Omega$cm/K$^{5/3}$ and 
$A_2 \approx 0.003\mu\Omega$cm/K$^2$. We will discuss interpretations of this behavior in Secs.~\ref{sec:4} and \ref{sec:5}, where we will
show that an equally good fit of the data is obtained by a superposition of $s=3/2$ and $s=2$.
\vskip -0cm
\begin{figure}[t]
\centering
\includegraphics[width=6 cm]{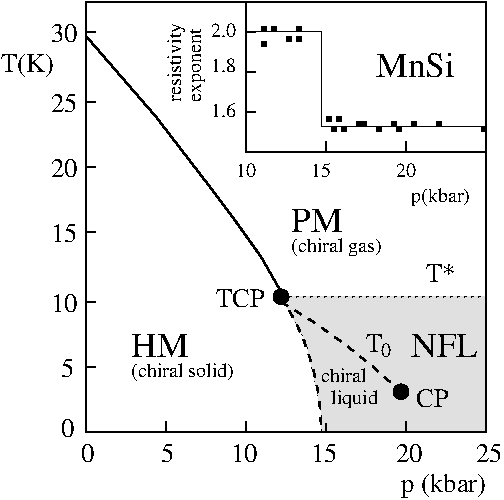}
\raisebox{-2mm}
{\hbox{
\includegraphics[width=9.1 cm]{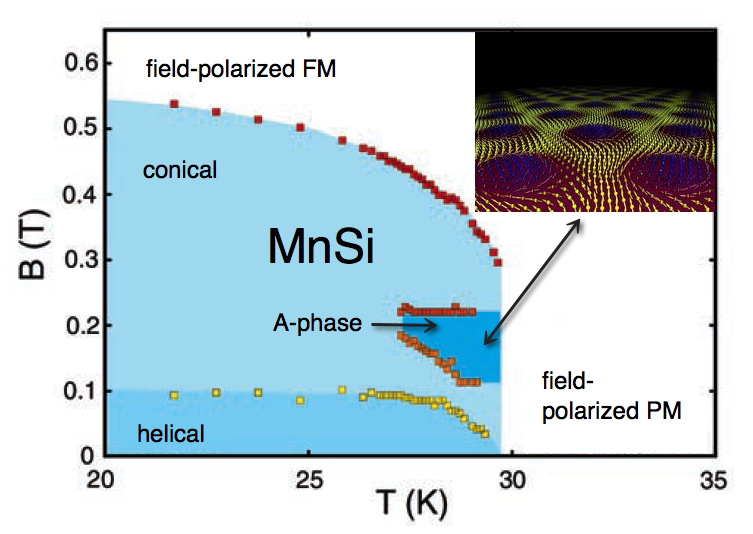}}
}
\caption{{\it Left panel:} Temperature-pressure phase diagram of MnSi based on experimental data from Refs.~\cite{Pfleiderer_et_al_1997, Pfleiderer_2007, 
              Pfleiderer_et_al_2004} and theoretical interpretations from Ref.~\cite{Tewari_Belitz_Kirkpatrick_2006}. HM and PM denote
              the helimagnetic and paramagnetic phases, respectively, and NFL denotes the region where $s=1.5$. The upper limit of
              the NFL region is not sharp. TCP is the observed tricritical point that separates second-order HM-PM transitions (solid line)
              from first-order ones (dashed line), and CP is a critical point proposed in Ref.~\cite{Tewari_Belitz_Kirkpatrick_2006}. The
              inset shows the abrupt change of the resistivity exponent at the critical pressure. From Ref.~\cite{Kirkpatrick_Belitz_2018}.
              {\it Right panel:} Magnetic field - temperature phase diagram of MnSi, showing various phases. The A-phase hosts a
              columnar skyrmionic spin texture. The inset shows an artist's rendition of the skyrmions in a plane perpendicular to the columns. 
              After Fig.~1 in Ref.~\cite{Muhlbauer_et_al_2009} (main figure), inset from Ref.~\cite{TUM_2009}.}
\label{fig:2}
\end{figure} 
\vskip 0cm
MnSi is a helimagnet with a rather long helical pitch wavelength of about 180\AA\ \cite{Ishikawa_et_al_1976}. Hydrostatic pressure destroys 
the helical order \cite{Pfleiderer_et_al_1997} and drives the system into a phase with no long-range magnetic order. There is, however, 
evidence for strong fluctuations in the nonmagnetic phase \cite{Pfleiderer_et_al_2004}. The phase diagram in the temperature-pressure 
plane is shown in the left panel of Fig.~\ref{fig:2}. The magnetic phase transition at low temperatures is first order, as is generically the case 
in clean metallic ferromagnets and long-wavelength helimagnets; for a review of the magnetic
properties see Ref.~\cite{Brando_et_al_2016}. Throughout the nonmagnetic phase, from the critical pressure $p_c\approx 15$kbar to
about 50 kbar, and over a temperature range from a few mK to almost 10K, the electrical resistivity displays a $T^{3/2}$ behavior with
a prefactor ranging from $A_{3/2} \approx 0.1\mu\Omega$cm/K$^{3/2}$ at high pressure to $0.22\mu\Omega$cm/K$^{3/2}$ near the
critical pressure \cite{Pfleiderer_Julian_Lonzarich_2001, Pfleiderer_2007}, see Figs.~\ref{fig:1}, \ref{fig:2}. In the helical phase the
electrical resistivity shows a $T^2$ behavior with a prefactor $A_2 \approx 0.03\mu\Omega$cm/K$^2$ at ambient pressure that rises,
first gradually and eventually sharply, to $A_2 \approx 0.12\mu\Omega$cm/K$^2$ as the critical pressure is approached from below
\cite{Pfleiderer_2007}. We note that these prefactors are surprisingly large. They are larger by a factor of 10 compared to their
counterparts in ZrZn$_2$, and larger by many orders of magnitude compared to the Coulomb scattering contribution given by the
Drude formula in conjunction with the scattering rate in Eq.~(\ref{eq:13}). The $T^2$ behavior in the helical phase is thus 
as anomalous as the $T^{3/2}$ behavior in the disordered phase, even though the exponent value happens to coincide with the
one characteristic of ordinary Fermi-liquid behavior. 

In a magnetic field, MnSi has a phase known as the A-phase that consists of a skyrmionic spin texture with the cores of the
skyrmions forming a hexagonal lattice of columns in the material \cite{Muhlbauer_et_al_2009}, see the right panel in Fig.~\ref{fig:2}. The $T^{3/2}$
behavior of the resistivity in the paramagnetic phase persists in a nonzero field up to the crossover to the field-polarized 
ferromagnetic region, and neutron scattering has provided evidence for strong fluctuations in the paramagnetic phase \cite{Pfleiderer_et_al_2004}.

Neither in the case of ZrZn$_2$ nor in that of MnSi is there any reason to believe that either critical fluctuations or diffusive electron
dynamics lead to the observed anomalous transport behavior. The explanation thus must lie in generic excitations that are extraneous
to the conduction electrons. We will discuss proposals along these lines in Sec.~\ref{sec:4}.

\begin{figure}[h]
\centering
\includegraphics[width=6 cm]{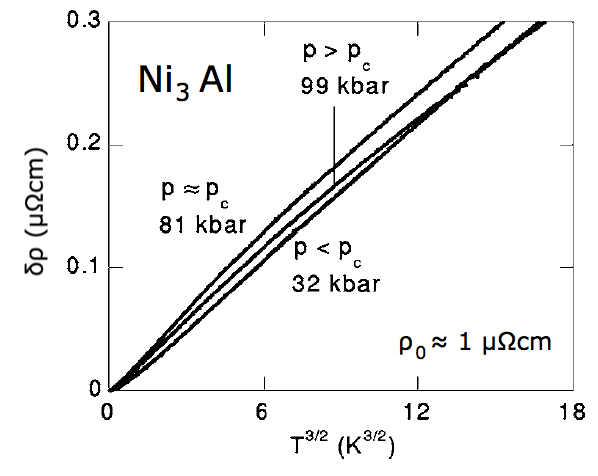}
\raisebox{4mm}
{\hbox{
\includegraphics[width=7.2cm]{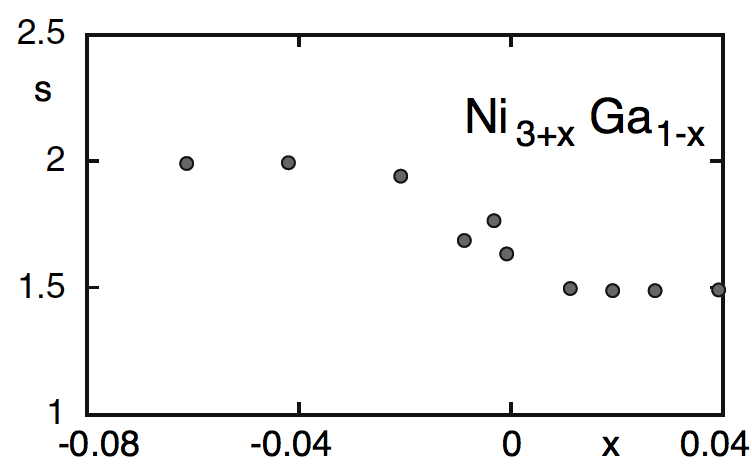}}
}
\caption{{\it Left panel:} Electrical resistivity of Ni$_3$Al plotted vs. $T^{3/2}$ for pressure values below, close to, and above the
              critical pressure $p_c$. After Fig.~4 in Ref.~\cite{Niklowitz_et_al_2005}. {\it Right panel:}  Resistivity exponent $s$ for Ni$_3$Al
              with a range of Ni concentrations. Data from Ref.~\cite{Fluitman_et_al_1973} as replotted in Ref.~\cite{Pfleiderer_2007}.
              After Fig.~3 in Ref.~\cite{Pfleiderer_2007}.
}
\label{fig:3}
\end{figure}

Another example of generic anomalous transport behavior in quantum magnets is provided by the isostructural compounds Ni$_3$Al and
Ni$_3$Ga, which can be prepared with various Ni concentration around the stoichiometric value. Ni$_3$Al shows ferromagnetic
order below 15-41K, depending on the exact composition, and a ferromagnetic-to-paramagnetic quantum phase transition (QPT) can
be triggered by means of hydrostatic pressure, see Refs.~\cite{Pfleiderer_2007, Brando_et_al_2016} and references therein. The transition is
suspected to be first order \cite{Niklowitz_et_al_2005, Pfleiderer_2007}, and stoichiometric samples have residual resistivities $\rho_0 \approx 1\mu\Omega$cm.
The resistivity exponent is $s\,{\substack{ >\\ \sim}}\,1.5$ on either side of the transition, see Fig.~\ref{fig:3}, and the prefactor
is $A_s \approx 0.01\mu\Omega$cm/K$^s$ \cite{Pfleiderer_2007}. Similar behavior is
observed in (Ni$_{1-x}$Pd$_x$)$_3$Al, which undergoes a ferromagnetic QPT at $x\approx 0.095$
\cite{Sato_1975}. This transport behavior is very similar to that observed in ZrZn$_2$. Stoichiometric Ni$_3$Ga is paramagnetic
and remains so for Ni-poor compositions, but it has a ferromagnetic ground state for Ni-rich compositions. In the ferromagnetic
samples $s \approx 1.5$ with a prefactor $A_{3/2}\approx 0.04\mu\Omega$cm/K$^{3/2}$, whereas $s=2$ in the paramagnetic
samples with $A_2\approx 0.001\mu\Omega$cm/K$^2$, see Ref.~ \cite{Pfleiderer_2007} and references therein. This situation is the reverse of the one in MnSi,
where $s=2$ in the magnetically ordered phase and $s=1.5$ in the disordered phase.
\vskip -0cm
\vskip 0cm
%

%%%%%%%%%%%%%%%%%%%%%%%%%%%%%%%%%%%%%%%%%%%%%%%%%%%%%%%%%%%%%%%%
\section{Theoretical explanations}
\label{sec:4}

As we have seen in Sec.~\ref{sec:2}, any explanation of the generic transport anomalies observed in various quantum magnets
must involve the scattering of electrons by soft generic excitations. In magnetically ordered phases, obvious candidates are the
Goldstone modes that result from the magnetic order. In phases without long-range magnetic order there are two possibilities.
Either, strong fluctuations that are remnants of the long-range order may provide scattering mechanisms that can lead to
generic transport anomalies. Candidates for such fluctuations have been observed in the nonmagnetic phase of MnSi
\cite{Pfleiderer_et_al_2004} and discussed as a possible origin of the observed $T^{3/2}$ behavior \cite{Kirkpatrick_Belitz_2010}.
Or, weak quenched disorder may provide droplets of the ordered phase within the disordered one (and vice versa) \cite{Kirkpatrick_Belitz_2016}.
This explains the widespread observation of phase separation away from the coexistence curve of a first-order of a first-order
phase transition, and it also provides  a way for scattering mechanisms that are germane to the magnetic phase to persist in
parts of the disordered phase. We will discuss these mechanisms in more detail later in this section, and also in Sec.~\ref{sec:5}.

For ferromagnets in the ordered phase, the magnon contribution to the resistivity was considered
early on and found to produce a $T^2$ behavior \cite{Kasuya_1956, Goodings_1963}, in agreement with experimental results
on Fe, Co, and Ni \cite{Campbell_Fert_1982}. Later work confirmed this, and also considered scattering by the continuum of
Stoner excitations (another example of the `unparticles' mentioned in Sec.~\ref{sec:2}) \cite{Ueda_Moriya_1975, Moriya_1985}.
The $T^2$ behavior is valid only above a characteristic temperature that is related to the exchange splitting \cite{Goedsche_Mobius_Richter_1979},
as we will see explicitly below. The behavior of both the electric and thermal resistivities in various temperature regimes was
discussed in Ref.~\cite{Bharadwaj_Belitz_Kirkpatrick_2014}. For helimagnets, the Goldstone modes and their
contribution to the scattering rates were derived in Refs.~\cite{Belitz_Kirkpatrick_Rosch_2006a, Belitz_Kirkpatrick_Rosch_2006b,
Kirkpatrick_Belitz_Saha_2008a, Kirkpatrick_Belitz_Saha_2008b}.
Recently, electron scattering from Goldstone modes in both ferromagnets and helimagnets has been reconsidered, and 
several new mechanism for anomalous transport behavior have been discussed \cite{Kirkpatrick_Belitz_2018}. In this section we give a
summary of the current state of the theory. We focus on three mechanisms that yield a resistivity exponent $s=3/2$ at least in some
temperature regime, for more complete results see Ref.~\cite{Kirkpatrick_Belitz_2018}.
For completeness, and to clarify some common misconceptions, we also briefly discuss the effects of non-generic critical fluctuations,
and the extent to which they exist.

\subsection{Scattering by magnetic Goldstone modes}
\label{subsec:4.1}

As we mentioned in Sec.~\ref{sec:2}, the basic considerations presented there need to be generalized and refined in
several ways in order to be applicable to magnetic materials. We start with a discussion of clean systems, and then consider
the effects of weak disorder.

\subsubsection{Clean systems}
\label{subsubsec:4.1.1}

In the ordered phase of both ferromagnets and helimagnets the effective local magnetic field seen by the conduction electrons
leads to an exchange splitting $\lambda$ of the Fermi surface. We thus need to distinguish between intraband scattering, where
a magnon is exchanged between electrons in the same sub-band of the exchange-split Fermi surface, and interband scattering,
where the exchange is between electrons in different sub-bands. At asymptotically low temperatures in clean systems the latter will always lead
to exponentially small rates, as the scattering processes get frozen out for temperatures small compared to the exchange
splitting. However, they can provide the leading contribution to scattering in a pre-asymptotic temperature window whose
lower limit can be rather low, and thus need to be considered. Furthermore, weak quenched disorder eliminates the exponential
suppression, as we will see. In ferromagnets, the magnons do not couple electrons in the
same sub-band, and thus interband scattering is the only mechanism available. The effective potential for interband scattering
is given by Eq.~(\ref{eq:7}), but with shifted arguments of the $\delta$-functions that reflect the fact that the electrons with
wave vector $\vec k$ live on a different Fermi surface than those with wave vector $\vec p$. The coupling constant $C$
is given by the square of the exchange interaction $\Gamma_t$, and the resulting expression for the single-particle 
interband scattering rate is
\be
\frac{1}{\tau_{\text{sp}}} \propto \NF\Gamma_t^2 \int_{-\infty}^{\infty} du\ \frac{1}{\sinh(u/T)} \frac{1}{\NF^2 V^2}\sum_{{\vec k},{\vec p}} \delta(\xi_{{\vec k}+{\vec p}}-\lambda)\,
        \delta(\xi_{\vec p}+\lambda)\, \chi''({\vec k},u)
\label{eq:14}
\ee
The transport rate is given by the same expression with an additional factor of ${\vec k}^2/\kF^2$ in the
integrand, with $\kF$ the Fermi wave number. This is known as the backscattering factor that suppresses large-angle scattering \cite{Wilson_1954}. 
For the transport interband scattering rate we thus have
\be
\frac{1}{\tau_{\text{tr}}} \propto \NF\Gamma_t^2 \int_{-\infty}^{\infty} du\ \frac{1}{\sinh(u/T)} \frac{1}{\NF^2 V^2}\sum_{{\vec k},{\vec p}} ({\vec k}^2/\kF^2)\,
        \delta(\xi_{{\vec k}+{\vec p}}-\lambda)\,\delta(\xi_{\vec p}+\lambda)\, \chi''({\vec k},u)
\label{eq:15}
\ee

\subsubsection{Systems with weak disorder}
\label{subsubsec:4.1.2}

Quenched disorder, however weak, is present in all real materials and leads to a nonzero scattering rate $1/\tau_0$ even at $T=0$
and a corresponding residual resistivity $\rho_0$. The cleanest samples of the magnetic systems discussed here have residual
resistivities of a few tenth of a $\mu\Omega$cm. While very clean by the standards of these compounds, these values are large
compared to the residual resistivities of many nonmagnetic metals (e.g., the residual resistivity of commercial Cu wire is less than 
$1\,$n$\Omega$cm). This motivates the consideration of disorder in the ballistic or weak-disorder regime \cite{Zala_Narozhny_Aleiner_2001}, 
which in magnets is characterized by the condition $\lambda\tau_0 \gg 1$ \cite{Kirkpatrick_Belitz_Saha_2008b}. A rigorous treatment requires 
elaborate diagrammatic calclations, but the net effect can be described by using simple heuristic arguments \cite{Kirkpatrick_Belitz_2018}.

Consider the expression for the clean single-particle rate in Eq.~(\ref{eq:14}). Performing the wave-number convolution integral yields
\be
\frac{1}{\NF V} \sum_{\vec p} \delta(\xi_{{\vec k}+{\vec p}}-\lambda)\,\delta(\xi_{\vec p}+\lambda) \propto \int_{-1}^1 d\eta\,\delta(k\vF\eta - 2\lambda)
   = \frac{1}{\vF\vert{\vec k}\vert}\,\Theta(\vert{\vec k}\vert - 2\lambda/\vF)
\label{eq:16}
\ee
where $\vF$ is the Fermi velocity. The step function leads to the exponential suppression of the rates at asymptotically low temperatures
mentioned above \cite{Goedsche_Mobius_Richter_1979, Bharadwaj_Belitz_Kirkpatrick_2014}. Weak disorder replaces the
$\delta$-function with a Lorentzian, and in the limit $\vF\vert{\vec k}\vert/\lambda \ll 1$, $\lambda\tau_0 \gg 1$ the step function gets
replaced by
\be
\frac{1}{\vF\vert{\vec k}\vert}\,\Theta(\vert{\vec k}\vert - 2\lambda/\vF) = \int_{-1}^1 d\eta\,\delta(k\vF\eta - 2\lambda) \rightarrow
   \int_{-1}^1 d\eta\,\frac{1/\tau_0}{(\vF\vert{\vec k}\vert\eta - 2\lambda)^2 + 1/\tau_0^2} \approx \frac{1}{\lambda^2\tau_0}
\label{eq:17}
\ee
The disorder thus eliminates the lower cutoff for the ${\vec k}$-integral and leads to an extra factor of $\vF\vert{\vec k}\vert/\lambda^2\tau_0$
in the integrand. Since $\vec k$ scales as a positive power of the temperature, this implies that the power-law $T$-dependence of the single-particle
rate is weaker than in the corresponding clean system, but extends to zero temperature. 

In the case of the transport rate, the same arguments apply, but in addition the disorder eliminates the backscattering factor since it leads
to more isotropic scattering. The effective extra factor in the integrand thus is $(\epsilonF/\lambda^2\tau_0)\kF/\vert{\vec k}\vert$, and the
disorder {\em strengthens} the $T$-dependence of the rate, in addition to eliminating the exponential suppression at asymptotically low $T$.
The single-particle rate and the transport rate thus are qualitatively the same and given by
\be
\frac{1}{\tau_{\text{sp}}} \propto \frac{1}{\tau_{\text{tr}}} \propto \frac{\NF\Gamma_t^2}{\lambda^2\tau_0} \int_{-\infty}^{\infty} du\,\frac{1}{\sinh(u/T)}\,\frac{1}{\NF V}\sum_{\vec k} \chi''({\vec k},u)
\label{eq:18}
\ee

$1/\tau_{\text{tr}}$ and $1/\tau_{\text{sp}}$ determine the electrical and thermal resistivities via the Drude formula
\be
\rho = m_{\text{e}}/n_{\text{e}}\, e^2 \tau
\label{eq:18'}
\ee
with $e$, $m_{\text{e}}$ and $n_{\text{e}}$ the electron charge, mass, and density, respectively.

\subsection{Application to quantum ferromagnets}
\label{subsec:4.2}

Now consider the scattering of electrons by magnons in ferromagnets. The relevant resonance frequency ($\omega_0$ in Sec.~\ref{sec:2}) is
\be
\omega_{\text{FM}}({\vec k}\to 0) = D{\vec k}^2
\label{eq:19}
\ee
with $D$ the spin-stiffness coefficient, and the corresponding susceptibility is, apart from a numerical prefactor \cite{Forster_1975},
\be
\chi_{\text{FM}}({\vec k},i\Omega) \propto \frac{m_0 D {\vec k}^2}{\omega_{\text{FM}}({\vec k})^2 - (i\Omega)^2}
\label{eq:20}
\ee
$m_0$ is the magnetization scale that determines the exchange splitting $\lambda$ via $\lambda = \Gamma_t m_0$. Two other
relevant energy scales are the largest energy that can be carried by a magnon (i.e., the magnetic equivalent of the Debye temperature)
\be
T_1 = D\kF^2
\label{eq:21}
\ee
and the smallest energy that can be transferred by means of magnon exchange,
\be
T_0 = D k_0^2 \approx T_1(\lambda/\epsilonF)^2
\label{eq:22}
\ee
where $k_0 = \lambda/\vF$.

For clean ferromagnets, Eqs.~(\ref{eq:14}, \ref{eq:15}) yield the results of Refs.~\cite{Bharadwaj_Belitz_Kirkpatrick_2014, Ueda_Moriya_1975, 
Goedsche_Mobius_Richter_1979}, viz., a $T\ln T$ and $T^2$ behavior for $1/\tau_{\text{sp}}$ and $1/\tau_{\text{tr}}$, respectively, for $T_0<T<T_1$,
and exponentially small rates for $T<T_0$. In the presence of ballistic quenched disorder, Eq.~(\ref{eq:18}) leads to both rates scaling as $T^{3/2}$,
which results in a resistivity contribution
\bse
\label{eqs:23}
\be
\delta\rho_{\text{FM}} = A_{3/2}^{\text{FM}}\, T^{3/2}
\label{eq:23a}
\ee
with a prefactor
\be
A_{3/2}^{\text{FM}} = \gamma_1\,\rho_0/T_1\sqrt{T_0}
\label{eq:23b}
\ee
\ese
where $\gamma_1$ is a numerical prefactor. Since the single-particle rate has the same $T$-dependence as the transport rate, this result also holds
for the thermal resistivity, only the numerical prefactor is different. It is valid for $T_{\text{ball}} \ll T\ \alt\ T_0$, with
\be
T_{\text{ball}} = T_1/(\epsilonF\tau_0)^2
\label{eq:24}
\ee
The lower limit on the temperature window is dictated by the constraints on the ballistic disorder regime \cite{Kirkpatrick_Belitz_2018}. For
lower temperatures, the electron dynamics are diffusive.

Parameter values appropriate for ZrZn$_2$ have been estimated in Ref.~\cite{Kirkpatrick_Belitz_2018}. The result is a prefactor
$A_{3/2}^{\text{FM}} \approx 0.01\mu\Omega$cm/K$^{3/2}$, which is very close to what is observed in this material, see the discussion in 
Secs.~\ref{sec:3} and \ref{sec:5}. In order for this mechanism to explain the anomalous transport behavior on either side of the first-order QPT, 
the droplet formation discussed in Ref.~\cite{Kirkpatrick_Belitz_2016} is crucial. 

\subsection{Application to quantum helimagnets}
\label{subsec:4.3}

In helimagnets there are two different soft modes that are candidates for explaining anomalous transport behavior. One are the helimagnons, which
are the Goldstone modes of the spontaneously broken symmetry that is present in the helically ordered phase \cite{Belitz_Kirkpatrick_Rosch_2006a}.
The other are fluctuations of the columnar skyrmion structure that is observed, e.g., in the A-phase of MnSi, see Fig.~\ref{fig:2} and Ref.~\cite{Muhlbauer_et_al_2009}.
Columnar fluctuations are familiar from the theory of liquid crystals \cite{DeGennes_Prost_1993} and have been studied in the context of helimagnets
in Refs.~\cite{Kirkpatrick_Belitz_2010} and \cite{Ho_et_al_2010}.

\subsubsection{Scattering by columnar fluctuations in skyrmionic phases}
\label{subsubsec:4.3.1}

Consider a hexagonal lattice of columns in the $z$-direction that fluctuate about their equilibrium positions, as shown in Fig.~4. Such fluctuations have an 
anisotropic dispersion relation, with the resonance frequency scaling linearly with the wave number for wave vectors ${\vec k}_{\perp}$ perpendicular to the 
\begin{figure}[t]
\centering
\includegraphics[width=4 cm]{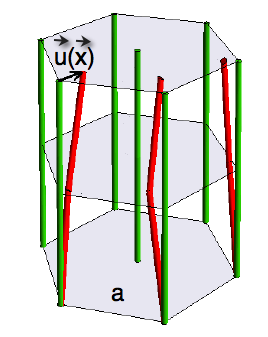}
\caption{Hexagonal lattice of columns, and fluctuations about this state. $a$ is the lattice constant, and ${\vec u}({\vec x})$ is the
              displacement vector. From Ref.~\cite{Ho_et_al_2010}.}
\label{fig:4}
\end{figure} 
%\vskip 0cm
%%%%%%%%%%
%\par\noindent
%\raisebox{2.1cm}
%{\parbox{10.0cm}{\baselineskip=13.0pt\sloppy
%   \vskip 1mm  
%         \par
%\medskip
%\hrule
%\medskip
%   {\rm {\bf Figure 4:} Hexagonal lattice of columns, and fluctuations about this state. $a$ is the lattice constant, and ${\vec u}({\vec x})$ is the
%              displacement vector. From Ref.~\cite{Ho_et_al_2010}.}
%   \medskip
%   \hrule
%   \vskip 2cm
%}  }
%\hfill
%\includegraphics[width=4.0cm]{fig_columnar_fluctuations.png}
%\vskip 5mm 
%\noindent
%%%%%%%%%%
columns, and quadratic with the wave number for wave vectors in the direction of the columns \cite{DeGennes_Prost_1993}. If the columnar structure is due
to skyrmions comprised by a superposition of three helices with pitch wave number $q$, as proposed in Ref.~\cite{Muhlbauer_et_al_2009}, then the
resonance frequency is \cite{Kirkpatrick_Belitz_2018, Petrova_Tchernyshyov_2011}
\be
\omega_{\text{sky}}({\vec k}) = \begin{cases} D\sqrt{k_z^4 + {\vec k}_{\perp}^2 q^2}\hskip 49pt \text{for}\quad D q^4/\kF^2\ \alt\ \omega_{\text{sky}} \alt\ D q^2 \\
                                                                        D(\kF^2/q^2)(k_z^4/q^2 + {\vec k}_{\perp}^2) \quad\text{for}\quad \omega_{\text{sky}}\ \alt\ D q^4/\kF^2
                                                 \end{cases}
\label{eq:25}
\ee
For $\omega_{\text{sky}}\ \agt\ D q^2$ the behavior crosses over to the ferromagnetic one given by Eq.~(\ref{eq:19}). The corresponding 
susceptibility is
\be
\chi_{\text{sky}}({\vec k},i\Omega) \propto \frac{m_0}{\omega_{\text{sky}}^2({\vec k}) - (i\Omega)^2}\begin{cases} 
          D q^2\hskip 70pt \text{for}\quad \omega_{\text{sky}}\ \agt\ D q^4/\kF^2 \\
          (\kF/q)^2 \omega_{\text{sky}}({\vec k})\qquad \text{for}\quad \omega_{\text{sky}}\ \alt\ D q^4/\kF^2
                    \end{cases}
\label{eq:26}
\ee
In the presence of ballistic disorder, the behavior of the mode in the upper frequency range leads, in conjunction with Eq.~(\ref{eq:18}), to the qualitatively same 
result for both the electrical and the thermal resistivity as in the ferromagnetic case,
\bse
\label{eqs:27}
\be
\delta\rho_{\text{sky}} = A_{3/2}^{\text{sky}}\,T^{3/2}
\label{eq:27a}
\ee
with a prefactor
\be
A_{3/2}^{\text{sky}} = \gamma_2\,\rho_0/T_1\sqrt{T_0}
\label{eq:27b}
\ee
\ese
where $\gamma_2$ is another numerical factor. This is valid for Max($T_{\text{ball}},T_q q^2/\kF^2)\ \alt\ T\ \alt\ T_q$, with
\be
T_q = D q^2
\label{eq:28}
\ee
another energy scale, this one relevant for helimagnets. Since the behavior for $T\ \agt\ T_q$ crosses over to the ferromagnetic one,
which is the same except for the numerical prefactor, the effective upper limit of the region of validity is the greater of $T_0$ and $T_q$.

For temperatures lower than $T_q(q/\kF)^2$ the lower frequency regime in Eqs.~(\ref{eq:25}, \ref{eq:26}) is relevant, and the $T$-dependence of
the resistivity crosses over to a $T^{5/4}$ behavior. However, for helimagnets with a small $q/\kF$ this crossover temperature
is extremely low and may not be larger than $T_{\text{ball}}$, so this behavior may not be observable. For instance, an estimate for
MnSi yields \cite{Kirkpatrick_Belitz_2018} $T_q \approx 250$mK, $T_{\text{ball}} \approx 1$mK. With $q/\kF\approx 0.03$ \cite{Ishikawa_et_al_1976},
this yields a crossover temperature of about 0.2 mK, which is lower than $T_{\text{ball}}$ and hence not observable.

With parameter values as appropriate for MnSi, an estimate of the prefactor shows that it is within a factor of 5 within what is observed in
the nonmagnetic phase of MnSi \cite{Kirkpatrick_Belitz_2018}. In order for this mechanism to be operative in that phase there must exist
strong columnar fluctuations. There is experimental evidence for this to be the case \cite{Pfleiderer_et_al_2004}. A theoretical analysis of
the possible nature of this phase was given in Ref.~\cite{Tewari_Belitz_Kirkpatrick_2006}.

\subsubsection{Scattering by helimagnons}
\label{subsubsec:4.3.2}

A third mechanism for a $T^{3/2}$ behavior of the electrical resistivity is provided by scattering of electrons by helimagnons, the Goldstone
modes of helical order, in clean helimagnets. The dispersion relation and the susceptibility for the helimagnons are \cite{Belitz_Kirkpatrick_Rosch_2006a}
\be
\omega_{\text{HM}}({\vec k}) = D \sqrt{q^2 k_z^2 + {\vec k}_{\perp}^4}
\label{eq:29}
\ee
and
\be
\chi_{\text{HM}}({\vec k},i\Omega) \propto \frac{m_0 D q^2}{\omega_{\text{HM}}^2({\vec k}) - (i\Omega)^2}
\label{eq:30}
\ee
Here $q$ is the modulus of the helical pitch wave vector, which we again take to point in the $z$-direction. This is valid for $\omega_{\text{HM}}\ \alt\ D q^2$, 
for larger wave numbers the behavior crosses over to the ferromagnetic one. We note that the numerator of the susceptibility is independent
of the wave number, whereas in the ferromagnet it is proportional to ${\vec k}^2$. As a result, the helimagnon susceptibility is softer than
the ferromagnon one, even though the Goldstone mode is stiffer in the helimagnet than in the ferromagnet. 

Equation~(\ref{eq:15}) now yields a contribution to the electrical resistivity that is given by \cite{Kirkpatrick_Belitz_2018}
\bse
\label{eqs:31}
\be
\delta\rho_{\text{HM}} = A_{3/2}^{\text{HM}}\,T^{3/2}
\label{eq:31a}
\ee
with
\be
A_{3/2}^{\text{HM}} = \rho_{\lambda} \gamma_3\,q/\kF T_1^{3/2}
\label{eq:31b}
\ee
\ese
Here $\rho_{\lambda} = \lambda m_{\text{e}}/n_{\text{e}} e^2$ is a resistivity scale, and $\gamma_3$ is another numerical factor.
This result is valid for $T_0\ \alt\ T\ \alt\ T_q$, provided this window exists; for $T\ \alt\ T_0$ the rate is exponentially suppressed.
In MnSi the window does not exist since $T_q < T_0$ \cite{Kirkpatrick_Belitz_2018}. 

In the presence of ballistic disorder, Eqs.~(\ref{eq:29}, \ref{eq:30}) in conjunction with Eq.~(\ref{eq:18}) yields a stronger $T$-dependence,
viz., $\delta\rho_{\text{HM}} \propto T\ln T$. At sufficiently low temperatures the observed $T^2$ behavior of unknown origin is predicted to
cross over to this behavior, see Ref.~\cite{Kirkpatrick_Belitz_2018} and the discussion in Sec.~\ref{sec:5} below.

\subsection{Scattering by critical fluctuations}
\label{subsec:4.4}

By now it is well established, both theoretically and experimentally, that the QPT in clean metallic
ferromagnets is generically first order \cite{Brando_et_al_2016}. However, historically  it was believed that
the transition is second order and hence is accompanied by critical fluctuations \cite{Hertz_1975}. We briefly discuss
the influence of these fluctuations on the resistivity, for two reasons: (1) The transition can be weakly first order in some materials,
and critical fluctuations may be observable in a pre-asymptotic regime, and (2) early work concerning the critical behavior
has influenced the analysis of experiments even in cases where later studies found a clear first-order transition.

Mathon \cite{Mathon_1968} found a resistivity exponent $s=5/3$ due to ferromagnetic quantum critical fluctuations even before Hertz \cite{Hertz_1975}
developed a renormalization-group treatment of QPTs. The same result was obtained by Millis \cite{Millis_1993} with
renormalization-group techniques; for a discussion of how this fits into a general scaling description of QPTs see Ref.~\cite{Kirkpatrick_Belitz_2015}.
An exponent $5/3 \approx 1.67$ can be experimentally hard to distinguish from 3/2, especially if there are various competing power-law contributions
to the resistivity that hold only in temperature windows of limited sizes, see Fig.~\ref{fig:5} and the related discussion. Furthermore, the critical fluctuations, if any, will be present only in a rather limited
region of the phase diagram and cannot explain observations of anomalous transport behavior far from any phase transition. Still, they may well 
contribute in parts of the phase diagram in materials where the QPT is weakly first order.

Sufficiently strong quenched disorder (strong enough to lead to diffusive electron dynamics) changes the nature of the ferromagnetic
QPT from first to second order. This was predicted theoretically \cite{Kirkpatrick_Belitz_1996, Belitz_et_al_2001b, Kirkpatrick_Belitz_2014}
and recently observed in Fe-doped MnSi \cite{Goko_et_al_2017}. The asymptotic critical behavior at this quantum critical point is unusual
and very hard to observe \cite{Belitz_et_al_2001b}, but in a pre-asymptotic region an effective power-law behavior with an exponent $s=3/11$
has been predicted \cite{Kirkpatrick_Belitz_2014, Kirkpatrick_Belitz_2015}.

\section{Discussion}
\label{sec:5}

We conclude by discussing various additional points and open problems. 

\subsection{Fermi liquids and non-Fermi liquids}
\label{subsec:5.1}

The anomalous transport behavior we have discussed in this paper is often referred to as non-Fermi-liquid (NFL) behavior. However, this term
has multiple meanings. Originally devised to describe the low-temperature behavior of fermions with a short-ranged interaction, such as 
He$^3$ \cite{Landau_1956, Landau_1957, Landau_1958}, Landau's Fermi-liquid theory was generalized to electrons with a long-ranged Coulomb 
interaction by Silin \cite{Silin_1957}. The chief concept of Fermi-liquid theory is the existence of quasiparticles that are continuously related to the
single-particle excitations in a Fermi gas. Accordingly, the term NFL is often used to refer to systems where the interactions are so strong that
they destroy the Landau quasiparticles. Examples are the Luttinger liquid \cite{Schulz_1995}, and the marginal Fermi liquid \cite{Varma_et_al_1989}
where the destruction is only logarithmic. A more readily observable feature of a Fermi liquid is an electrical (and thermal) resistivity that has a
$T^2$ temperature dependence for $T\to 0$ due to Coulomb scattering, see Sec.~\ref{subsec:2.2}. (We have, however, glossed over various
complications, see Ref.~\cite{Pal_Yudson_Maslo_2012}.) Systems where this is not the case are
also often referred to as NFLs. However, it is important to keep in mind that NFL transport behavior in this sense does not imply that no
Landau quasiparticles exist, it may merely mean that there are soft excitations that scatter the conduction electrons more strongly than the
Coulomb interaction does. We have discussed three examples of such excitations that are generic, namely, ferromagnons, columnar fluctuations, and 
helimagnons, and one that requires fine tuning, namely, ferromagnetic critical fluctuations.

\subsection{Mechanism for generic scale invariance}
\label{subsec:5.2}

There are a limited number of mechanisms that lead to generic soft modes, and generic scale invariance, in many-particle systems. 
Three common ones are: (1) Spontaneously broken continuous symmetries that lead to Goldstone modes, (2) conservation laws, and 
(3) gauge symmetries, see Ref.~\cite{Belitz_Kirkpatrick_Vojta_2005} for a comprehensive discussion. The three examples we have discussed 
all belong to the first category. They all are two-particle excitations, i.e., correlation functions of four fermion fields. In clean fermion systems the 
single-particle excitations described by the Green function are also soft. References~\cite{Belitz_Kirkpatrick_Vojta_2005, Brando_et_al_2016} 
also discussed how rare regions in systems with quenched disorder fit into the classification scheme of generic scale invariance. This scarcity of 
generic soft modes, especially ones that can lead to a linear $T$-dependence of the electrical resistivity, is part of the motivation for suggestions 
that a hidden quantum critical point underlies the ``strange-metal'' normal state of high-T$_{\text{c}}$ superconductors (for a discussion see, e.g., 
\cite{Sachdev_2010}). There is currently no consensus on the origin of this behavior. The phenomenological marginal-Fermi-liquid description of 
Ref.~\cite{Varma_et_al_1989} is a generic mechanism that yields a resistivity exponent $s=1$, but the microscopic origin of the marginal Fermi 
liquid is not clear. 

\subsection{Uniqueness, or lack thereof, of the resistivity exponent}
\label{subsec:5.3}

It is important to note that generically there are many competing contributions to the $T$-dependence of the resistivity, and usually
more than one are of comparable strength in any given temperature regime. Examples of a well-defined exponent $s$ over a sizable 
temperature range, such as $s=3/2$ in MnSi, or $s=1$ in high-T$_{\text{c}}$ superconductors, are rare and suggest one strongly dominant
scattering mechanism in these systems. More commonly, the value of $s$ is less well defined and changes as a function of $T$,
see the experimental data for ZrZn$_2$ in Fig.~\ref{fig:1} and Ni$_3$Al in Fig.~\ref{fig:3}. Qualitatively, this is easy to understand
from a slight extension of the discussion we have given in Sec.~\ref{sec:4}. We have focussed on scattering mechanisms that
result in $s=3/2$, however, a more complete analysis shows that there are various mechanisms in various temperature windows
that lead to values of $s$ between $s=1$ and $s=2$, see Table I in Ref.~\cite{Kirkpatrick_Belitz_2018}.

%
%\vskip -0cm
\begin{figure}[t]
\centering
\includegraphics[width=12 cm]{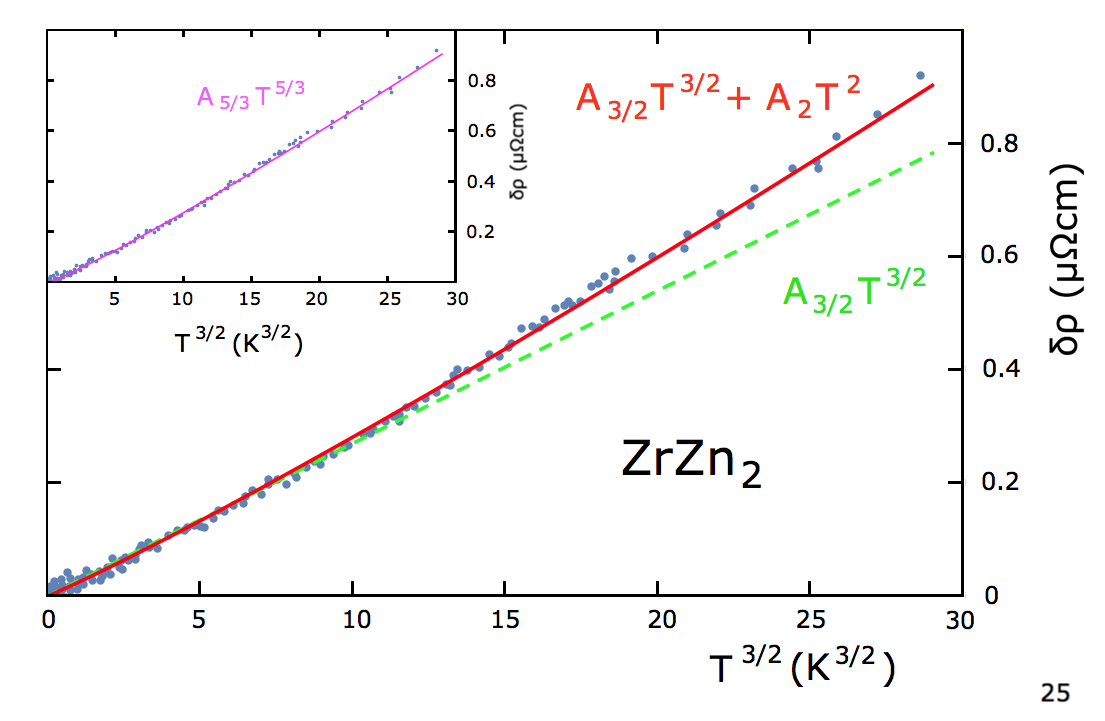}
\caption{Resistivity data (blue dots) of ZrZn$_2$ vs. $T^{3/2}$ at ambient pressure. Data (blue dots) from Fig. 4 of Ref.~\cite{Takashima_et_al_2007}.
              The solid red line is a fit using Eq.~(\ref{eq:5.1}) with $s=3/2$, $A_{3/2} = 0.021\,\mu\Omega$cm/K$^{3/2}$ and $A_2 = 0.0033\,\mu\Omega$cm/K$^2$.
              The dashed green line is a pure $T^{3/2}$ fit with $A_{3/2} = 0.027\,\mu\Omega$cm/K$^{3/2}$. The inset shows a pure a pure $T^{5/3}$ fit with
              $A_{5/3} = 0.0215\,\mu\Omega$cm/K$^{5/3}$. On the scale of the figure, this fit is indistinguishable from the red line in the main figure.
              }
\label{fig:5}
\end{figure} 
%\vskip 0cm
%
To illustrate this point, let us discuss the behavior of ZrZn$_2$ at ambient pressure in more detail. References \cite{Takashima_et_al_2007}
and \cite{Sutherland_et_al_2012} found that the behavior of the electrical resistivity between 1K and about 15K is well described by Eq.~(\ref{eq:1})
with $s=5/3$ and $A_{5/3} \approx 0.02\ \mu\Omega$cm/K$^{5/3}$. However, in general one would always expect a $T^2$ contribution (of
Fermi-liquid origin or otherwise) that is additive to the leading contribution with $s<2$. One should thus write
\be
\delta\rho(T\to 0) = A_s T^s + A_2 T^2 + o(T^2)
\label{eq:5.1}
\ee
References.~\cite{Takashima_et_al_2007, Sutherland_et_al_2012} used $s$ and $A_s$ as given above, and $A_2=0$, and obtained a good fit.
In Fig.~\ref{fig:5} we reproduce this fit (in the inset) and compare it with a fit that uses Eq.~(\ref{eq:5.1}) with $s=3/2$, $A_{3/2} = 0.021\,\mu\Omega$cm/K$^{3/2}$, 
and $A_2 = 0.0033\,\mu\Omega$cm/K$^2$ (solid red line in the main figure). There are at least two physical motivations for this: (1) In Sec.~\ref{sec:4} we have identified several 
scattering mechanisms that lead to $s=3/2$. (2) There is no reason to believe that $A_2=0$. Indeed, Ref.~\cite{Sutherland_et_al_2012} found a
$T^2$ behavior in a magnetic field of 9T with a prefactor that is very close to the one used for the fit in Fig.~\ref{fig:5}. An obvious explanation is
that the magnetic field gaps out the magnons, which eliminates the scattering mechanism that produces $s=3/2$, and leaves a $T^2$ mechanism
of unknown origin behind. It then is natural to assume that this $T^2$ mechanism is also present in zero field and needs to be taken into account.
It is very interesting that the resulting fit, the solid red line in the main figure, is of equal quality as the pure $T^{5/3}$ fit shown in the inset. Indeed,
if plotted on top of one another the two fits are indistinguishable on the scale of the figure. We conclude that the data by themselves cannot distinguish
between a pure $T^{5/3}$ behavior and a $T^{3/2}$ behavior with a $T^2$ correction. A pure $T^{3/2}$ behavior, on the other hand, gives a good fit
only in a much more limited temperature regime, see the green dashed line in the figure.

In this context of multiple scattering mechanisms we also stress again that a resistivity exponent $s=2$ does not necessarily imply that the transport behavior is
conventional. For instance, the very large value of the prefactor $A_2$ observed in the helically ordered phase of MnSi cannot
be explained by any known scattering mechanism. A related point is that a scattering mechanism leading to a smaller value of $s$
may not dominate over one leading to a larger value unless one goes to very low temperatures, as the crossover temperature
obvious depends on the ratio of the prefactors. In the helical phase of MnSi, the helimagnons (in the form of the $T\ln T$ contribution
mentioned at the end of Sec.~\ref{subsubsec:4.3.2}) must manifest themselves at
sufficiently low temperatures, but an estimate in Ref.~\cite{Kirkpatrick_Belitz_2018} suggests that they will dominate over the
unknown scattering mechanism leading to $s=2$ only for temperatures below about 30 mK. Transport measurements in the
ordered phase of MnSi to check this prediction would be very interesting.

\subsection{Quenched disorder}
\label{subsec:5.4}

An important component of the discussion of systems with weak quenched disorder in Sec.~\ref{subsubsec:4.1.2} was that the disorder
suppresses the backscattering factor in the expression for the transport relaxation rate. This is a qualitative argument, and a more detailed
theoretical analysis of the disorder dependence of the backscattering factor, and the related crossover in the $T$-dependence of the electrical
resistivity, is desirable. The same is true for the crossover from the ballistic or weak-disorder regime to the strong-disorder regime that is
characterized by diffusive electron dynamics. On the experimental side, a more detailed characterization of disorder, and how to quantify its
presence, would be desirable. The residual resistivity may be a rather crude measure of disorder. For instance, there is experimental 
evidence for inhomogeneities in pressure-tuned systems that are not necessarily reflected in transport experiments and thus can be present
even in systems with a rather small residual resistivity \cite{Yu_et_al_2004}.

%%%%%%%%%%%%%%%%%%%%%%%%%%%%%%%%%%%%%%%%%%
%\section{Patents}
%This section is not mandatory, but may be added if there are patents resulting from the work reported in this manuscript.

%%%%%%%%%%%%%%%%%%%%%%%%%%%%%%%%%%%%%%%%%%
\vspace{6pt} 

%%%%%%%%%%%%%%%%%%%%%%%%%%%%%%%%%%%%%%%%%%
%% optional
%\supplementary{The following are available online at \linksupplementary{s1}, Figure S1: title, Table S1: title, Video S1: title.}

% \supplementary{The following are available at \linksupplementary, Figure S1: title, Table S1: title, Video S1: title. A supporting video article is available at doi: link.}

%%%%%%%%%%%%%%%%%%%%%%%%%%%%%%%%%%%%%%%%%%
\authorcontributions{Both authors have contributed equally to all aspects of this study.}

%%%%%%%%%%%%%%%%%%%%%%%%%%%%%%%%%%%%%%%%%%
\funding{This work was supported by the NSF under grant numbers DMR-1401449 and DMR-1401410. Part of this work was performed at the
              Aspen Center for Physics, supported by the NSF under Grant No. PHYS-1066293.}

%%%%%%%%%%%%%%%%%%%%%%%%%%%%%%%%%%%%%%%%%%
\acknowledgments{We thank Achim Rosch, Ronojoy Saha, and Sripoorna Bharadwaj for collaborations, and Andrey Chubukov and Arnulf M{\"o}bius for discussions.}

%%%%%%%%%%%%%%%%%%%%%%%%%%%%%%%%%%%%%%%%%%
\conflictsofinterest{The authors declare no conflict of interest.} 

%%%%%%%%%%%%%%%%%%%%%%%%%%%%%%%%%%%%%%%%%%
%% optional
\abbreviations{The following abbreviations are used in this manuscript:\\

\noindent 
\begin{tabular}{@{}ll}
NFL & Non-Fermi Liquid \\
QPT & Quantum Phase Transition \\
\end{tabular}}

%%%%%%%%%%%%%%%%%%%%%%%%%%%%%%%%%%%%%%%%%%
%% optional
\appendixtitles{no} %Leave argument "no" if all appendix headings stay EMPTY (then no dot is printed after "Appendix A"). If the appendix sections contain a heading then change the argument to "yes".
\appendixsections{multiple} %Leave argument "multiple" if there are multiple sections. Then a counter is printed ("Appendix A"). If there is only one appendix section then change the argument to "one" and no counter is printed ("Appendix").
%\appendix
%\section{}
%\subsection{}
%The appendix is an optional section that can contain details and data supplemental to the main text. For example, explanations of experimental details that would disrupt the flow of the main text, but nonetheless remain crucial to understanding and reproducing the research shown; figures of replicates for experiments of which representative data is shown in the main text can be added here if brief, or as Supplementary data. Mathematical proofs of results not central to the paper can be added as an appendix.

%\section{}
%All appendix sections must be cited in the main text. In the appendixes, Figures, Tables, etc. should be labeled starting with `A', e.g., Figure A1, Figure A2, etc. 


%=====================================
% References, variant A: internal bibliography
%=====================================
\reftitle{References}

% The following MDPI journals use author-date citation: Arts, Econometrics, Economies, Genealogy, Humanities, IJFS, JRFM, Laws, Religions, Risks, Social Sciences. For those journals, please follow the formatting guidelines on http://www.mdpi.com/authors/references
% To cite two works by the same author: \citeauthor{ref-journal-1a} (\citeyear{ref-journal-1a}, \citeyear{ref-journal-1b}). This produces: Whittaker (1967, 1975)
% To cite two works by the same author with specific pages: \citeauthor{ref-journal-3a} (\citeyear{ref-journal-3a}, p. 328; \citeyear{ref-journal-3b}, p.475). This produces: Wong (1999, p. 328; 2000, p. 475)

%=====================================
% References, variant B: external bibliography
%=====================================
%\externalbibliography{yes}
%\bibliography{your_external_BibTeX_file}

\begin{thebibliography}{999}
%
%[1]
\bibitem{Wilson_1954}
Wilson, A.~H. 
The theory of metals. 
Cambridge University Press: Cambridge, UK, 1954.
%
%[2]
\bibitem{Kittel_2005}
Kittel, C. 
Introduction to Solid State Physics. 
Wiley: New York, 2005.
%
%[3]
\bibitem{Gurvitch_Fiory_1989}
Gurvitch, M., and Fiory, A.~T.
Resistivity of La$_{1.825}$Sr$_{0.175}$CuO$_4$ and YBa$_2$Cu$_3$O$_7$ to 1100K: Absence of Saturation and Its Implications
{\em Phys. Rev. Lett.} {\bf 1989}, {\em 59}, 1337.
%
%[3a]
\bibitem{Takagi_et_al_1992}
Takagi, H., Batlogg, B., Kao, H.~L., Kwo, J., Cava, R.~J., Krajewski, J.~J., and Peck, W.~F.
Systematic Evolution of Temperature-Dependent Resistivity in La$_{2-x}$Sr$_x$CuO$_4$
{\em Phys. Rev. Lett.} {\bf 1992}, {\em 69}, 2975.
%
%[4]
\bibitem{Sato_1975}
Sato, M. 
Magnetic Properties and Electrical Resistivity of (Ni$_{1-x}$Pd$_x$)$_3$Al. 
{\em J. Phys. Soc. Japan} {\bf 1975}, {\em 39}, 98.
%
%[5]
\bibitem{Niklowitz_et_al_2005}
Niklowitz, P.~G., Beckers, F., Lonzarich, G.~G., Knebel, G., Salce, B., Thomasson, J., Bernhoeft, N., Braithwaite, D., and Flouquet, J. 
Spin-fluctuation-dominated electrical transport of Ni$_3$Al at high pressure.
{\em Phys. Rev. B} {\bf 2005}, {\em 72}, 024424. 
%
%[6]
\bibitem{Takashima_et_al_2007}
Takashima, S., Nohara, M., Ueda, H., Takeshita, N., Terakura, C., Sakai, F., and Takagi, H.
Robustness of the Non-Fermi-Liquid Behavior near the Ferromagnetic Critical Point in Clean ZrZn$_2$.
{\em J. Phys. Soc. Japan} {\bf 2007}, {\em 76}, 043704.
%
%[7]
\bibitem{Pfleiderer_Julian_Lonzarich_2001}
Pfleiderer, C., Julian, S.R., and Lonzarich, G.~G.
Non-Fermi-liquid nature of the normal state of itinerant-electron ferromagnets
{\em Nature} {\bf 2001}, {\em 414}, 427.
%
%[8]
\bibitem{Kirkpatrick_Belitz_2018}
Kirkpatrick, T.~R., and Belitz, D.
Generic non-Fermi-liquid behavior of the resistivity in weakly disordered ferromagnets and clean helimagnets
{\em Phys. Rev. B} {\bf 2018}, {\em 97}, 064411.
%
%[9]
\bibitem{Forster_1975}
Forster, D. {\em Hydrodynamic Fluctuations, Broken Symmetry, and Correlation Functions}; Benjamin: Reading, MA, 1975.
%
%[10]
\bibitem{Ziman_1960}
Ziman, J.~M.
{\em Electrons and Phonons};
Clarendon: Oxford, UK, 1960.
%
%[11]
\bibitem{Pines_Nozieres_1989}
Pines, D. and Nozi{\`e}res, P.
{\em The Theory of Quantum Liquids};
Addison-Wesley: Redwood City, CA, 1989.
%
%[12]
\bibitem{Georgi_2007}
Georgi, H.
Unparticle Physics
{\em Phys. Rev. Lett.} {\bf 2007}, {\em 98}, 221601.
%
%[13]
\bibitem{Anderson_1984}
Anderson, P.W.
{\em Basic Notions of Condensed Matter Physics}; Benjamin: Menlo Park, CA, 1984; ch. 3.A.
%
%[14]
\bibitem{Bharadwaj_Belitz_Kirkpatrick_2014}
Bharadwaj, S., Belitz, D., and Kirkpatrick, T.~R.
Electronic relaxation rates in metallic ferromagnets.
{\em Phys. Rev. B} {\bf 2014}, {\em 84}, 134401.
%
%[15]
\bibitem{Brando_et_al_2008}
Brando, M., Duncan, W.~J., Moroni-Klementowicz, D., Albrecht, C., Gr{\"u}ner, D., Ballou, R., and Grosche, F.~M.
Logarithmic Fermi-Liquid Breakdown in NbFe$_2$.
{\em Phys. Rev. Lett.} {\bf 2008}, {\em 101}, 126401.
%[16]
\bibitem{Lee_Ramakrishnan_1985}
Lee, P.~A. and Ramakrishnan, T.~V.
Disordered electronic systems.
{\em Rev. Mod. Phys.} {\bf 1985}, {\em 57}, 287.
%
%[17]
\bibitem{Belitz_Kirkpatrick_1994}
Belitz, D. and Kirkpatrick, T.~R.
The Anderson-Mott transition.
{\em Rev. Mod. Phys.} {\bf 1994}, {\em 66}, 261.
%
%[17a]
\bibitem{Uhlarz_Pfleiderer_Hayden_2004}
Uhlarz, M., Pfleiderer, C., and Hayden, S.~M.
Quantum Phase Transition in the Itinerant Ferromagnet ZrZn$_2$
{\em Phys. Rev. Lett.} {\bf 2004}, {\em 93}, 256404.
%
%[17b]
\bibitem{Pfleiderer_et_al_1997}
Pfleiderer, C., McMullan, G.~J., Julian, S.~R., and Lonzarich, G.~G.
Magnetic quantum phase transition in MnSi under hydrostatic pressure.
{\em Phys. Rev. B} {\bf 1997}, {\em 55}, 8330.
%
%[18]
\bibitem{Brando_et_al_2016}
Brando, M., Belitz, D., Grosche, F.~M., and Kirkpatrick, T.~R.
Metallic quantum ferromagnets.
{\em Rev. Mod. Phys} {\bf 2016}, {\em 88}, 025006.
%
%[19]
\bibitem{Sutherland_et_al_2012}
Sutherland, M., Smith, R.~P., Marcano, N., Zou, Y., Rowley, S.~E., Grosche, F.~M., Kimura, N., Hayden, S.~M., Takashima, S., Nohara, M., and Takagi, H.
Transport and thermodynamic evidence for a marginal Fermi-liquid state in ZrZn$_2$.
{\em Phys. Rev. B} {\bf 2012}, {\em 85}, 035118.
%
%[19a]
\bibitem{Ishikawa_et_al_1976}
Ishikawa, Y., Tajima, K., Bloch, D., and Roth, M.
Helical spin structure in manganese silicide MnSi
{\em Solid State Commun.} {\bf 1976}, {\em 19}, 525.
%
%[19b]
\bibitem{Pfleiderer_et_al_2004}
Pfleiderer, C., Reznik, D., Pintschovius, L., {v.~L{\"o}hneysen}, H., Garst, M., and Rosch, A.
Partial order in the non-Fermi-liquid phase of MnSi.
{\em Nature} {\bf 2004}, {\em 427}, 227.
%
%[20]
\bibitem{Pfleiderer_2007}
Pfleiderer, C.
On the Identification of Fermi-Liquid Behavior in Simple Transition Metal Compounds.
{\em J. Low Temp. Phys.} {\bf 2007}, {\em 147}, 231.
%
%[20a]
\bibitem{Tewari_Belitz_Kirkpatrick_2006}
Tewari, S., Belitz, D., and Kirkpatrick, T.~R.
Blue Quantum Fog: Chiral Condensation in Quantum Helimagnets.
{\em Phys. Rev. Lett.} {\bf 2006}, {\em 96}, 047207.
%
%[21]
\bibitem{Muhlbauer_et_al_2009}
M{\"u}hlbauer, S., Binz, B., Jonietz, F., Pfleiderer, C., Rosch, A., Neubauer, A., Georgii, R., and B{\"o}ni, P.
Skyrmion Lattice in a Chiral Magnet.
{\em Science} {\bf 2009}, {\em 323}, 915.
%
%[22]
\bibitem{TUM_2009}
Technical University Munich Press Release.
Discovery of a new magnetic order: Skyrmion Lattice in a Chiral Magnet.
https://www.frm2.tum.de/en/news-media/press/news/news/article/discovery-of-a-new- magnetic-order-skyrmion-lattice-in-a-chiral-magnet/
({\bf 2009}).
%
%[27]
\bibitem{Fluitman_et_al_1973}
Fluitman, J.~H.~J., Boom, R., {De~Chatel}, P.~F., Schinkel, C.~J., Tilanus,J.~L.~L., and {De~Vries}, B.~R.
Possible explanations fo the low temperature resisitivities of Ni$_3$Al and Ni$_3$Ga alloys in terms of spin density fluctuation theories
{\em J. Phys. F} {\bf 1973}, {\em 3}, 109.
%
%[28]
\bibitem{Kirkpatrick_Belitz_2010}
Kirkpatrick, T.~R., and Belitz, D.
Columnar Fluctuations as a Source of Non-Fermi-Liquid Behavior in Weak Metallic Magnets.
{\em Phys. Rev. Lett.} {\bf 2010}, {\em 104}, 256404.
%
%[29]
\bibitem{Kirkpatrick_Belitz_2016}
Kirkpatrick, T.~R., and Belitz, D.
Stable phase separation and heterogeneity away from the coexistence curve.
{\em Phys. Rev. B} {\bf 2016}, {\em 93}, 144203.
%
%[30]
\bibitem{Kasuya_1956}
Kasuya, T.
Electrical Resistance of Ferromagnetic Metals.
{\em Progr. Theor. Phys.} {\bf 1956}, {\em 16}, 58.
%
%[31]
\bibitem{Goodings_1963}
Goodings, D.~A.
Electrical Resistivity of Ferromagnetic Metals at Low Temyeratures.
{\em Phys. Rev.} {\bf 1963}, {\em 132}, 542.
%
%[32]
\bibitem{Campbell_Fert_1982}
Campbell, I.~A. and Fert, A.
Transport Properties of Ferromagnets. In {\em Ferromagnetic Materials, Vol. 3}; Wohlfarth, E.~P., Ed; North Holland: Amsterdam, 1982
%
%[33]
\bibitem{Ueda_Moriya_1975}
Ueda, K. and Moriya, T.
Contribution of Spin Fluctuations to the Electrical and Thermal Resistivities of Weakly and Nearly Ferromagnetic Metals.
{\em J. Phys. Soc. Japan} {\bf 1975}, {\em 39}, 605.
%
%[34]
\bibitem{Moriya_1985}
Moriya, T.
{\em Spin Fluctuations in Itinerant Electron Magnetism}; Springer: Berlin, 1985.
%
%[35]
\bibitem{Goedsche_Mobius_Richter_1979}
Goedsche, F., M{\"o}bius, A., and Richter, A.
On the Low-Temperature Resistivity of Ferromagnetic Transition Metal Alloys.
{\em phys. stat. sol. (b)} {\bf 1979}, {\em 96}, 279.
%
%[36]
\bibitem{Belitz_Kirkpatrick_Rosch_2006a}
Belitz, D., Kirkpatrick, T.~R., and Rosch, A.
Theory of helimagnons in itinerant quantum systems.
{\em Phys. Rev. B} {\bf 2006}, {\em 73}, 054431.
%
%[37]
\bibitem{Belitz_Kirkpatrick_Rosch_2006b}
Belitz, D., Kirkpatrick, T.~R., and Rosch, A.
Theory of helimagnons in itinerant quantum systems. II. Nonanalytic corrections to Fermi-liquid behavior.
{\em Phys. Rev. B} {\bf 2006}, {\em 74}, 094408.
%
%[38]
\bibitem{Kirkpatrick_Belitz_Saha_2008a}
Belitz, D., Kirkptrick, T.~R., and Saha, R.
Theory of helimagnons in itinerant quantum systems. III. Quasiparticle description.
{\em Phys. Rev. B} {\bf 2008}, {\em 78}, 094407.
%
%[39]
\bibitem{Kirkpatrick_Belitz_Saha_2008b}
Belitz, D., Kirkptrick, T.~R., and Saha, R.
Theory of helimagnons in itinerant quantum systems. IV. Transport in the weak-disorder regime.
{\em Phys. Rev. B} {\bf 2008}, {\em 78}, 094408.
%
%[40]
\bibitem{Zala_Narozhny_Aleiner_2001}
Zala, G., Narozhny, B.~N., and Aleiner, I.~L.
Interaction corrections at intermediate temperatures: Longitudinal conductivity and kinetic equation.
{\em Phys. Rev. B} {\bf 2001}, {\em 64}, 214204.
%
%[41]
\bibitem{DeGennes_Prost_1993}
DeGennes, P.~G., and Prost, J.
{\em The Physics of Liquid Crystals};
Clarendon: Oxford, UK, 1993.
%
%[42]
\bibitem{Ho_et_al_2010}
Ho, {K-y.}, Kirkpatrick, T.~R., Sang, Y., and Belitz, D.
Ordered Phases of Itinerant Dzyaloshinsky-Moriya Magnets and Their Electronic Properties.
{\em Phys. Rev. B} {\bf 2010}, {\em 82}, 134427.
%
%[43]
\bibitem{Petrova_Tchernyshyov_2011}
Petrova, O., and Tchernyshyov, O.
Spin waves in a skyrmion crystal
{\em Phys. Rev. B} {\bf 2011}, {\em 84}, 214433.
%
%[44]
\bibitem{Mathon_1968}
Mathon, J.
Magnetic and Electrical Properties of Ferromagnetic Alloys Near the Critical Concentration
{\em Proc. Roy. Soc. London, Series A} {\bf 1968}, {\em 306}, 355.
%
%[45}
\bibitem{Hertz_1975}
Hertz, J.
Quantum Phase Transitions.
{\em Phys. Rev. B} {\bf 1975}, {\em 14}, 1165.
%
%[46]
\bibitem{Millis_1993}
Millis, A.~J.
Effect of a nonzero temperature on quantum critical points in itinerant fermion systems
{\em Phys. Rev. B} {\bf 1993}, {\em 48}, 7183.
%
%[47]
\bibitem{Kirkpatrick_Belitz_2015}
Kirkpatrick, T.~R., and Belitz, D.
Exponent relations at quantum phase transitions, with applications to metallic quantum ferromagnets.
{\em Phys. Rev. B} {\bf 2015}, {\em 91}, 214407.
%
%[48}
\bibitem{Kirkpatrick_Belitz_1996}
Kirkpatrick, T.~R., and Belitz, D.
Quantum critical behavior of disordered itinerant ferromagnets.
{\em Phys. Rev. B} {\bf 1996}, {\em 53}, 14364.
%
%[49]
\bibitem{Belitz_et_al_2001b}
Belitz, D., Kirkpatrick, T.~R., Mercaldo, M.~T., and Sessions, S.
Quantum critical behavior in disordered itinerant ferromagnets: Logarithmic corrections to scaling
{\em Phys. Rev. B} {\bf 2001}, {\em 63}, 174428.
%
%[50]
\bibitem{Kirkpatrick_Belitz_2014}
Kirkpatrick, T.~R., and Belitz, D.
Pre-asymptotic critical behavior and effective exponents in disordered metallic quantum ferromagnets
{\em Phys. Rev. Lett.} {\bf 2014}, {\em 113}, 127203.
%
%
%[51]
\bibitem{Goko_et_al_2017}
Goko, T., Arguello, C.~J., Hamann, A., Wolf, T., Lee, M., Reznik, D., Maisuradze, A., Khasanov, R., Morenzoni, E., and Uemura, Y.~T.
Restoration of quantum critical behavior by disorder in pressure-tuned (Mn,Fe)Si
{\em npj Quantum Materials} {\bf 2017}, {\em 2},44.
%
%[53]
\bibitem[Landau(1956)]{Landau_1956}
Landau, L.~D. 
The theory of a Fermi liquid. 
{\em Zh. Eksp. Teor. Fiz.} {\bf 1956}, {\em 30}, 1058 [Sov. Phys. JETP {\bf 1957}, {\em 3}, 920].
%
%[54]
\bibitem[Landau(1957)]{Landau_1957}
Landau, L.~D. 
Oscillations in a Fermi liquid. 
{\em Zh. Eksp. Teor. Fiz.} {\bf 1957}, {\em 32}, 59 [Sov. Phys. JETP {\bf 1957}, {\em 5}, 101].
%
%[55]
\bibitem[Landau(1958)]{Landau_1958}
Landau, L.~D. On the theory of the Fermi liquid. 
{\em Zh. Eksp. Teor. Fiz.} {\bf 1958}, {\em 35}, 97 [Sov. Phys. JETP {\bf 1959}, {\em 8}, 70].
%
%[56]
\bibitem[Silin(1957)]{Silin_1957}
Silin, V.~P. 
Theory of a degenerate electron liquid. 
{\em Zh. Eksp. Teor. Fiz.} {\bf 1957}, {\em 33}, 459 [Sov. Phys. JETP {\bf 1958}, {\em 6}, 387].
%
%[57]
\bibitem{Schulz_1995}
Schulz, H.
Fermi liquids and non-Fermi liquids.
In {\em Proceedings of the Les Houches Summer School LXI}; Akkermans, E., Montambqux, G., Pichard, J., and Zinn-Justin, J., Eds.; Elsevier: Amsterdam, 1995.
%
%[58]
\bibitem{Varma_et_al_1989}
Varma, C.~M., Littlewood, P.~B., Schmitt-Rink, S., Abrahams, E., and Ruckenstein, A.~E.
Phenomenology of the Normal State of Cu-O High-Temperature Superconductors.
{\em Phys. Rev. Lett.} {\bf 1989}, {\em 63}, 1996.
%
%[59]
\bibitem{Pal_Yudson_Maslo_2012}
Pal, H.~K., Yudson, V.~I., and Maslov, D.~L.
Resistivity of Non-Galilean-Invariant Fermi- and Non-Fermi Liquids
{\em Lith. J. Phys.}, {\bf 2012}, {em 52}, 142.
%
%[60]
\bibitem{Belitz_Kirkpatrick_Vojta_2005}
Belitz, D., Kirkpatrick, T.~R., and Vojta, T.
How generic scale invariance influences quantum and classical phase transitions.
{\em Rev. Mod. Phys.} {\bf 2005}, {\em 77}, 579.
%
%[61]
\bibitem{Sachdev_2010}
Sachdev, S.
Where is the quantum critical point in the cuprate superconductors?
{\em Phys. Status Solidi B.} {\bf 2010}, {\em 247}, 537.
%
%[62]
\bibitem{Yu_et_al_2004}
Yu, W., Zamborsky, F., Thompson, J.~D., Sarrao, J.~L., Torelli, M.~E., Fisk, Z., and Brown, S.~E.
{\em Phys. Rev. Lett.} {\bf 2004}, {\em 92}, 086403.
%
%\bibitem{Ong_1990} 
%The Hall Effect and its Relation to other Transport Phenomena in the Normal State of the High-Temperature Superconductors.
%In {\em Physical Properties of High Temperature Superconductors II}; Ginsberg, D.~M., Ed; World Scientific: Singapore, 1990.
%
% Reference 1
%\bibitem[Author1(year)]{ref-journal}
%Author1, T. The title of the cited article. {\em Journal Abbreviation} {\bf 2008}, {\em 10}, 142-149, doi:xxxxx.
% Reference 2
%\bibitem[Author2(year)]{ref-book}
%Author2, L. The title of the cited contribution. In {\em The Book Title}; Editor1, F., Editor2, A., Eds.; Publishing House: City, Country, 2007; pp. 32-58, ISBN.
\end{thebibliography}

%%%%%%%%%%%%%%%%%%%%%%%%%%%%%%%%%%%%%%%%%%
%% optional
%\sampleavailability{Samples of the compounds ...... are available from the authors.}

%% for journal Sci
%\reviewreports{\\
%Reviewer 1 comments and authors’ response\\
%Reviewer 2 comments and authors’ response\\
%Reviewer 3 comments and authors’ response
%}

%%%%%%%%%%%%%%%%%%%%%%%%%%%%%%%%%%%%%%%%%%
\end{document}